# Data Ethics in the Era of Healthcare Artificial Intelligence in Africa: An Ubuntu Philosophy Perspective

Abdoul Jalil Djiberou Mahamadou[a*], Aloysius Ochasi[b], Russ B. Altman[c, d, e, f]

[a]Stanford Center for Biomedical Ethics, Stanford University, Stanford, CA 94305, USA, abdjiber@stanford.edu; *Corresponding Author
[b]Department of Bioethics and Interdisciplinary Studies, Brody School of Medicine, East Carolina University, Greenville, NC 27858, USA
[c]Department of Biomedical Data Science, Stanford University, Stanford, CA 94305, USA
[d]Department of Bioengineering, Stanford University, Stanford, CA 94305, USA
[e]Department of Genetics, Stanford University, Stanford, CA 94305, USA
[f]Department of Medicine, Stanford University, Stanford, CA 94305, USA

## Abstract

Data are essential in developing healthcare artificial intelligence (AI) systems. However, patient data collection, access, and use raise ethical concerns, including informed consent, data bias, data protection and privacy, data ownership, and benefit sharing. Various ethical frameworks have been proposed to ensure the ethical use of healthcare data and AI, however, these frameworks often align with Western cultural values, social norms, and institutional contexts emphasizing individual autonomy and well-being. Ethical guidelines must reflect political and cultural settings to account for cultural diversity, inclusivity, and historical factors such as colonialism. Thus, this paper discusses healthcare data ethics in the AI era in Africa from the Ubuntu philosophy perspective. It focuses on the contrast between individualistic and communitarian approaches to data ethics. The proposed framework could inform stakeholders, including AI developers, healthcare providers, the public, and policy-makers about healthcare data ethical usage in AI in Africa.

**Keywords**: data ethics, artificial intelligence, ubuntu philosophy, ethical framework, global health

## Introduction

Healthcare systems are the pillar of public health and well-being, providing essential services to communities worldwide. However, only between one-third and one-half of the world's population had access to essential health services in 2017 (World Health Organization 2020), especially in the Global South. Although countries experience similar healthcare access barriers, they are more pronounced in the Global South (Dawkins et al. 2021). Main access barriers include sociocultural, demographic, patient, financial, environmental, infrastructure, and treatment, which delay the decision to seek care, reach an adequate facility, and receive care once at the facility (Dawkins et al. 2021). In Africa, poor health outcomes have been associated with inadequate human resources, budget allocation, and poor management and leadership (Oleribe et al. 2019). Improving population health and lessening the healthcare divide requires implementing appropriate healthcare reforms, which can be time-consuming and costly. As a remedy, digital technologies, particularly Artificial Intelligence (AI), offer promising, efficient, and cost-effective solutions (Khanna et al. 2022).

AI is "an umbrella term for a range of techniques that can be used to make machines complete tasks in a way that would be considered intelligent were they to be completed by a human" (Morley et al. 2020). In healthcare, in addition to the reduction of healthcare expenditures, between $200 billion and $360 billion in 2019 in the US (Sahni et al. 2023), AI has been successfully deployed for treatment recommendation, patient care, and administrative activities (Davenport and Kalakota 2019). Medical diagnosis is a notable application of AI, where deep learning attained and outperformed human experts (Bajwa et al. 2021; Liu et al. 2019). While nascent in the Global South, AI has played a substantial role in disease outbreak predictions and surveillance, assessing morbidity and mortality risks, and supporting health policy and planning design (Schwalbe and Wahl 2020). For instance, in Africa, with the use of AI and climate, environmental, meteorological, and geospatial data, significant progress has been made toward improving disease outbreak predictions and forecasting of diarrhea, typhoid, anemia, Ebola, tuberculosis, malaria, HIV/AIDS, Zika, and COVID-19 (Mbunge and Batani 2023). This results in minimizing disease transmission and spread, fostering behavioral change, and adequate resource allocations (Mbunge and Batani 2023). Also, AI has the potential to aid global healthcare institutions in tackling the public health crisis of an aging population and the rising prevalence of chronic illnesses and cancers exerting immense pressure on an already strained global health system, especially in the Global South (Roppelt, Kanbach, and Kraus 2024).

Despite the tremendous potential to improve health outcomes and equity in the Global South, the development of healthcare AI raises a wide range of concerns among others resulting from data collection, storage, and sharing (Moodley and Rennie 2023). These concerns include informed consent, data privacy, and protection, data ownership, data sharing, and profit, and data bias. Furthermore, healthcare data processing by AI systems introduces additional concerns such as transparency, accountability, and social justice at different model development stages (Char, Abràmoff, and Feudtner 2020; Martinez-Martin et al. 2021). Ethical, legal, and regulatory AI and data governance frameworks have been proposed to minimize AI risks and maximize its benefits (Micheli et al. 2020; Floridi and Cowls 2019; Cath 2018; Al-Badi, Tarhini, and Khan 2018). However, the ethical AI debate is dominated by Western countries (Jobin, Ienca, and Vayena 2019) and reflects Western cultural values, social norms, and institutional contexts (Ayana et al. 2023). The dominance of Western ideals and standards in AI negates that value systems differ across cultures. In essence, certain matters have high moral significance or are considered critical problems in some societies that may not be considered so in other societies (Birhane 2020). For instance, moral judgment in mainstream ethical frameworks prioritizes individuals or their properties such as autonomy and well-being (Wareham 2017). Under the mainstream perspective, personal relationships such as familial or communal have little moral significance, however, these relationships have moral primacy in African cultures, especially in Ubuntu philosophy (Wareham

2017). This tension between individuality and community moral judgment has several implications including the applicability and effectiveness of mainstream ethical frameworks in Africa and the need to contextualize these frameworks to the African cultural, political, and economic settings.

In this paper, we discuss healthcare data ethics in the era of AI in Africa from the perspective of Ubuntu philosophy. Ubuntu is an African normative theory in which "actions are morally right to the extent that they honour the capacity to relate communally, reduce discord or promote friendly relationships with others, and in which the physical world and the spiritual world are fundamentally united" (Ewuoso and Hall 2019). It represents a general notion of universal interdependence, solidarity, and communalism. Ubuntu advocates for human interactions that nurture sharing and building trusting relationships filled with mutual compassion, respect, listening, and affirming others (Nussbaum 2003; Gilliam 2021). Ubuntu philosophy encompasses various aspects, due to the limited scope of this paper, we will focus solely on its communal characteristics (Ewuoso and Hall 2019). We argue for a cultural and context-specific view of data ethics by addressing data colonialism, informed consent, data availability and quality effect on AI appropriateness, data privacy, and protection, data ownership and profit, data sharing, and data bias.

1. Data Colonialism

Colonialism is "a practice of domination, which involves the subjugation of one people to another" (Kohn and Reddy 2006). In the digital World, colonialism may take the form of data colonialism defined as "the decentralized extraction and control of data from citizens with or without their explicit consent 'through communication networks developed and owned by Western tech companies'" (Coleman 2019). Closely allied with data colonialism is algorithmic colonialism, which is driven by profit maximization and the desire to "dominate, monitor, and influence social, political, and cultural discourse through the control of core communication and infrastructure mediums" (Birhane 2020). While traditional colonialism championed by mostly European governments employed brute force as a tool of domination over the colonized people, data colonialism is spearheaded by corporate tech monopolies often masquerading as 'digital saviors' helping to liberate an impoverished continent. In an era of AI when data revolution in Africa is described as the 'new gold' or the 'new oil' or even 'data-rich continent' (Moodley and Rennie 2023), Western tech companies motivated by profit could perceive Africans as "human natural resources" free for the taking (Zuboff, 2019). Such assumptions and ideological domination fail to realize that 'data are people' who should not be objectified, exploited, or manipulated for corporate profit. Data colonialism dehumanizes people. Dehumanization refers to "stripping people of human qualities" where persons are viewed more as objects rather than as equally human moral agents (Bandura 1999).

While tech monopolies and data colonialism are not issues specific to Africa and the Global South, low-resource countries are more vulnerable to their effects due to their historical colonial context, resulting in an imbalance of power between the Global North and Global South and technological dependence. Furthermore, low-resource countries are less equipped to address the challenges posed by data colonialism. In high-resource countries, regulatory bodies such as the EU General Data Protection Regulation (GDPR) and the US Health Insurance Portability and Accountability Act (HIPPA) are safeguards for controlling the type of data that can be collected and analyzed. Moreover, Global North countries have the power to regulate tech companies to limit their influence which little or non-Global South countries have. Safeguard initiatives such as the GDPR and HIPPA especially in the context of AI are nonexistent or slow to develop in Africa (ALT Advisory 2022).

Besides the implications previously discussed, data colonialism can undermine trust between the Global North and Global South, particularly in data collection in the Global South. Data from the Global South are essential to balance populations' representativeness in biomedical data. However, the fear of data colonialism from Global North companies can prevent these populations from giving their data. This could lead to AI technologies suited for only certain populations and missed opportunities to benefit from AI in the Global South.

Morally, Western tech companies involved in the development and implementation of healthcare AI in Africa should not be driven by profit maximization at any cost; rather, they consider the overall well-being of individuals and communities, especially the vulnerable, and not impoverish the "development of local products" or "leaving the continent dependent on its software and infrastructure" (Birhane 2020). Granting a high moral consideration to these issues will aid in the responsible mining of data to develop AI that would benefit Africa and the Global South.

2. **Informed consent**

Informed consent has become an integral component of modern medicine, respecting the patient's right to self-determination (autonomy) in decisions regarding their medical care and participation in clinical trials. It requires the "disclosure of all necessary information that a reasonable person would use in making an informed decision, in a format that is readily understandable to the individual, and without coercion influencing the choice" (B. A. Townsend and Scott 2019; European Health Telematics Association 2012). Informed consent is anchored on the Western notion of autonomy. Whereas this idea of autonomy is overtly individualistic, sub-Saharan thinking tends to be more communitarian. Communitarianism is a concept based on the idea that the aspirations and actions of the individual are constitutive and an extension of the community of which she is part (Waghid 2013). Mbiti, a prominent cultural anthropologist, articulated the communitarian nature of African culture thus: "In African culture, only in terms of other people does the individual become conscious of his own being, his own duties, his privileges and responsibilities towards himself and towards other people" (Mbiti 1990). Individuals in an African communitarian milieu describe their self-identity by the maxim 'I am, because we are, and since we are, therefore I am' (Mbiti 1990). This view contrasts with the Western individualistic culture, where people describe their autonomous self-identification by the Cartesian maxim, Cogito ergo sum (I think, therefore I am). Given the differences in both cultures, informed consent rooted in individualism may not have the same currency in a more communitarian setting. For instance, in the Western traditional articulation of informed consent, the voluntary consent of a research subject or legally designated surrogates is required for participation in research, but in an African communitarian setting, community consent given by community leaders, village heads, or elders may be sufficient for research since "in some settings, the notion of individual informed consent can seem inappropriate, because important decisions are often made in conjunction with families or are even left to communities" (Sugarman et al. 2001; Frimpong-Mansoh 2008). While moral considerations should be given to cultural values, care must be taken to protect every research participant from unnecessary harm.

The informed consent process becomes more complex with the development and deployment of healthcare AI. The two aspects of the informed consent process in healthcare AI are informed consent for data use and informed consent in clinical encounters.

   a. *Informed consent for data use*

Big data, described as 'the oxygen on which AI depends,' is integral to building algorithms and machine learning (Petrozzino 2021). The big data revolution in Africa is promoted as having potential benefits for healthcare but also raises ethical concerns about how the data is collected, stored, and shared (Moodley and Rennie 2023). When companies collect, process, and store patients' data, especially personally identifiable information, it is ethically obligatory to obtain informed consent. Personal information is broadly defined as "information that identifies, relates to, describes, is reasonably capable of being associated with, or could reasonably be linked, directly or indirectly, with a particular consumer or household," and personal data is defined as "any information relating to an identified or identifiable natural person" (Gerke, Minssen, and Cohen 2020). The explicit consent of every individual is needed if their personally identifiable information or personal data will be used to train AI algorithms for clinical use or if their data is shared (Kim and Routledge 2022; Siala and Wang 2022). Patients should be informed about the collection and use of their data and be given meaningful choices "to opt in or opt out of specific data sharing or analysis activities, based on their preferences and values" (Godwin Olaoye and Luz 2024). Failure to secure informed consent violates autonomy unless consent is waived for ethically acceptable reasons. As indicated earlier, in an African communitarian culture, community consent is given by community leaders, village heads, or elders for research in addition to the individual consent as an extra layer of protection geared towards ensuring that the individual participants are not exposed to unnecessary harm (nonmaleficence).

   b. *Informed consent in clinical encounters*

The traditional informed consent process in medical care involves full disclosure, patient comprehension, and consent. Questions arise on how these elements are deployed in the clinical AI space, and how the healthcare practitioner notifies the patient about the complexities and involvement of AI in their care. The disclosure dilemma is further compounded by the "black-box" phenomenon where AI algorithms result from noninterpretable machine techniques. The inability of clinicians to explain how AI makes health recommendations raises questions about transparency, accountability, and the "right to explanation" (Felder 2021). Should doctors offer treatments that they do not understand? Does it undermine the patient's trust in physicians, which is necessary for the therapeutic relationship to flourish? The American Medical Association Journal of Ethics notes the difficulty that might arise in the informed consent process: "When an AI device is used, the presentation of information can be complicated by possible patient and physician fears, overconfidence, or confusion" (Schiff and Borenstein 2019; Gerke, Minssen, and Cohen 2020). The informed consent process is complicated in high-income countries where patient autonomy and the right to self-determination are the norm; it is even more complicated in low to middle-income countries where paternalism is still entrenched. The respect for patients' autonomy and human dignity demands that healthcare practitioners everywhere converse with patients about using AI in their clinical encounters. Every patient should be allowed to decline the use of AI in their evaluation and treatment. The rules of informed consent should be applied across the board in clinical AI; however, particular attention must be given to the African cultural context, which may be more communitarian than individualistic.

3. **Data Protection and Privacy**

Every individual has the right and desire to control the release or disclosure of personal health information. Any unauthorized disclosure of personal information and data violates a patient's privacy and erodes trust in the medical establishment. AI depends on sensitive and personal information to train algorithms, which raises many privacy challenges and security concerns. Some of the challenges include the unauthorized collection, use, and sharing of a patient's data, not providing patients with access and control over their data and how it is used, data triangulation and potential for reidentification of anonymized data, and data breaches by cybercriminals (Beverley Alice Townsend et al. 2023; Astromskė, Peičius, and Astromskis 2021; Murphy et al. 2021; Cohen et al. 2014). To ensure patient privacy-preserving when training AI systems, technical solutions including Differential Privacy (Dwork 2006) and Federated Learning (McMahan et al. 2017) have been proposed, however, these techniques often result in accuracy loss (Torkzadehmahani et al. 2022) which may raise additional ethical concerns.

The traditional Western formulation of privacy and confidentiality stipulates that patient have the right to privacy and expect medical professionals to keep information gathered in therapeutic encounters private and confidential. In an African communitarian milieu, the demands of privacy and confidentiality may have less moral significance than in the typical Western setting because of an individual's obligation to others, especially family members. Consequently, what the individual does with her body is only partially up to the individual in certain respects. The rationale is that "since other members of the community have a stake in the individual's health, many Africans would think that they ought to be aware of her illness and play a role in discussing how she ought to treat it" (Metz and Gaie 2010; Kasenene 2000). Ultimately, the emphasis is more on the community and not autonomy, but it does not mean the destruction of individuality. An average African might not prioritize autonomy if they realize that the common or collective good of the family and community is promoted. It is, therefore, a moral imperative that the development and implementation of clinical AI align with the values of African communitarian culture.

As healthcare providers create, receive, maintain, and transmit large amounts of sensitive and high-quality patient data necessary to optimize AI algorithms to their precision, particular attention should be paid to protecting it. When a patient's data is used outside of the doctor-patient relationship, it might negatively impact their health, insurance premiums, job opportunities, and personal relationships and even cause embarrassment. Such misuse violates the principle of nonmaleficence (Do no harm). These concerns are further exacerbated in Africa, where private companies and unscrupulous individuals may exploit the lack of legal protections to misuse patients' data (Kaur, Garg, and Gupta 2021). According to a 2022 study, 35% of African countries have no data protection legislation (ALT Advisory 2022). The UNCTAD also noted that 35% of LMICs have no data protection legislation, with 19% in Africa in 2021 (United Nations Conference on Trade and Development 2021). Governments and regulatory agencies must prioritize passing or strengthening data protection laws in Africa to assure the public that their data will be used fairly and adequately in AI research.

4. **Data Ownership and Profit**

Data ownership and profit sharing raise a plethora of legal and ethical challenges. The legal issues involved will not be discussed due to the limited scope of this paper. Generating large amounts of health data worth billions of dollars raises the question of ownership and who can access it for research or commercial purposes, primarily when generated through collaborative efforts or public data (Wahl et al. 2018). Some evidence suggests the public's discomfort with companies or governments selling patient data for profit (Gerke, Minssen, and Cohen 2020; Lords 2018). We argue that private and public institutions engaged in significant data generation in Africa should prioritize public or common interest over corporate profit. It is morally obligatory for all institutions to view AI data generated with public funds as a public good and ensure that the profits are used responsibly and reciprocally to provide tangible health benefits for all. This could take many forms, including establishing a repository for large amounts of readily available data accessible to researchers as a global public good (Wahl et al. 2018; Tindana et al. 2019). In addition to serving the public good, efforts should be made to ensure that individual patients whose data are used also benefit. It is laudable that in LMICs, "those seeking to use patient data must show that they are adding value to the health of the very same patients whose data is being used" (Cohen 2018).

5. **Data Bias**

AI bias occurs when AI systems favor certain individuals or groups over others resulting in unfair outcomes. Bias arising from the collection, use, and sharing of data often constitutes the main source of AI bias. This includes historical bias, aggregation bias, representation bias, and sampling bias (Mehrabi et al. 2021). In healthcare, when unaddressed, biased data can amplify and perpetuate health disparities (Celi et al., 2022) and socio-economic inequalities (Ade-Ibijola and Okonkwo 2023) which can further marginalize communities in low-resource countries, especially in Africa. Gender bias is another major concern in Africa where women and girls face persistent disparities in almost every aspect of life (Ahmed and Sey 2020). Thus, it becomes morally imperative to identify and mitigate bias and enforce fairness to minimize the translation of gender stereotypes and discrimination into AI systems (UNESCO 2021) and promote equitable representation and decision-making processes.

Despite the wide range of technical solutions to address data and AI bias (Barocas, Hardt, and Narayanan 2023), no consensus on effective approaches has emerged (Cary et al. 2023). Moreover, technical fairness enforcement solutions predominantly reflect Western distributive justice theories (Carey and Wu 2023; Gajane and Pechenizkiy 2017) which often conflict with African distributive justice (Wareham, 2019). With data scarcity in Africa, AI tools developed in the West and fine-tuned on African populations offer opportunities to address certain data and AI biases such as gender bias. Nevertheless, these tools may carry their own biases making their deployment ineffective in Africa or even adding new ethical concerns to the existing ones discussed in this paper. For instance, a fair model developed in some environmental settings (e.g., in the US) can be unfair in other environmental settings (e.g., Africa) due to dataset shifts (Finlayson Samuel G. et al. 2021).

6. Data Sharing

Health data-sharing processes in Africa should engage local communities, align with their values, norms, and culture, and address ethical concerns. It is well-recognized that sharing health data can enable scientific progress to improve population health. In AI, data sharing can foster transparency and address ethical concerns, including mitigating different types of AI bias (Norori et al. 2021; Gaonkar, Cook, and Macyszyn 2020). Nevertheless, appropriate data sharing, especially health research data, remains challenging in low-resource countries and raises ethical concerns (Bull, Roberts, and Parker 2015; Bull et al. 2015). Notable concerns include the protection of participants' privacy and confidentiality, consent validity, data ownership and control, moral distance, limited awareness of the context of the data collection, and impacts on public trust (Bull, Roberts, and Parker 2015). In Africa, power asymmetry, trust issues, benefit-sharing, and the lack of contextual knowledge of data collection exacerbate data-sharing challenges (Abebe et al. 2021). These interconnected challenges drive imbalances, inequalities, and injustices in the data-sharing ecosystem (Abebe et al. 2021). Furthermore, beyond financial and infrastructure burdens, the political sphere in Africa, shaped by the legacies of colonialism and the prioritization of Western-centric needs and values in data-sharing, adds another layer of complexity (Abebe et al. 2021). Other factors include concerns about shared data misuse, misunderstanding, and misleading secondary data analysis results (Denny et al. 2015).

Along with the ethical concerns, several factors, including motivational, technical, economic, and legal, may infringe on data sharing in low-resource countries (Schwalbe et al. 2020). A recent comparative study of health data sharing governance in five African countries, for instance, shows that inconsistency in health data definitions, lack of cross-border data transfer guidelines, and lack of data subject participation in data sharing processes constitute the primary barriers to effective data sharing (Nienaber McKay et al. 2024).

## 7. Data Availability, Quality and AI Appropriateness

Healthcare AI in Africa raises concerns about appropriateness. Appropriateness or suitability addresses how and why AI-driven technologies are being used (Fletcher, Nakeshimana, and Olubeko 2021). Data scarcity and low quality in Africa could hinder the development, implementation, and use of healthcare AI technologies by impacting the performance and generalizability of the models (Bansal, Sharma, and Kathuria 2022), making these technologies unsuitable for available data. Inappropriate use of AI in the healthcare setting could significantly affect patient outcomes. For instance, deploying AI models trained on adult data to pediatric data could result in substantive consequences (Muralidharan et al. 2023). Besides the performance and generalizability, inadequate data quality could amplify AI systems' bias, unfairness, and opacity (Vibbi 2024). Technical solutions such as data augmentation, transfer learning, and few-shots learning have been proposed to improve AI models' performance and robustness under low-data settings (Shorten and Khoshgoftaar 2019), however, these solutions could raise additional concerns such as dataset shift under which a model can be fair on one dataset and unfair on another (Barrainkua et al. 2022).

Evaluating the appropriateness of AI systems is critical to minimize harm. Numerous AI appropriateness frameworks have been proposed including the AI Suitability Framework (NetHope 2020) developed by NetHope, a consortium of sixty nonprofit organizations. NetHope identified thirty-two questions to guide AI practitioners in determining the appropriateness of AI. Under this framework, AI is evaluated along different dimensions, from defining the opportunity of using AI to appraising the data and bias and the resources needed to develop, implement, and maintain the technology. Initially intended for nonprofit organizations, the proposed framework can be adapted to AI for healthcare in Africa and provides an overview of the factors to consider when evaluating AI systems' appropriateness. Here, we emphasized the potential benefits of AI and the feasibility of implementing AI-driven solutions.

The first category focuses on measuring and evaluating AI's concrete advantages over traditional non-AI approaches. These advantages include healthcare efficiency improvement, decision-making capabilities enhancement, increased automation, handling complex tasks with greater accuracy and speed, and cost savings. In Africa, as discussed in the introduction, AI has enabled informed healthcare resource allocation and improved disease outbreak predictions and forecasting (Mbunge and Batani 2023). Nonetheless, various factors influence the feasibility of integrating these technologies into existing systems and workflows. These barriers encompass technical considerations such as data availability, quality, compatibility with AI algorithms, and infrastructure requirements such as computing power and storage capacity which are scarce in Africa. Additionally, organizational factors such as skills and expertise, regulatory compliance, cultural readiness, and the ethical considerations discussed in this work play crucial roles in determining the practicality of AI implementation. Therefore, evaluating AI's appropriateness requires an adequate risk and benefit analysis tailored to Africa's healthcare context and an analysis of the resources needed through the life cycle of AI technologies.

## 8. Limitations

We argued for a more communitarian perspective of mainstream ethical standards. We acknowledge the ongoing debate of whether social groups or communities should have moral rights, particularly in granting consent (Schrag 2006). In the West, group, and community consent are often viewed as safeguarding historically marginalized or abused groups and communities in medical research or required for community engagement. Under this setting, communities may have legal powers to control researchers and research protocols in the community. These rational and community characteristics contrast with the African context. Irrespective of historical research misconduct, African societies, particularly rural and low-literate settings, are highly influenced by cultural values and traditional practices (Appiah 2021), making communitarianism a norm rather than an exception to universal moral standards. Nevertheless, a communitarian approach to bioethics presents some limitations, specifically when individual and gatekeeper interests compete. We further acknowledge that Ubuntu is neither a representative African normative theory nor unanimously accepted across the continent (Ewuoso and Hall 2019).

**Conclusion**

We have provided an overview of challenges to ethical and responsible AI in healthcare that aligns with human values and leads to human flourishing in Africa. We argue that a context-specific and culturally appropriate ethical framework that respects African values and norms is integral to successfully deploying and implementing healthcare AI on the continent. Thus far, a wide range of ethical frameworks initiated to promote the responsible use of AI often fall under the common moral theories umbrella that reflect the idea of autonomy (self-determination) in Western societies. However, overtly individualistic autonomy does not have the same self-evidence in African contexts that tend to be more communitarian. AI deployment in the continent without ethical oversights can harm individuals. The real-life consequences of unchecked and unethical AI are enormous because they harm the end users who are unaware that what is touted as a blessing to Africa is stealthily destroying it. In his book 2084: Artificial Intelligence and the Future of Humanity, the Oxford mathematician and philosopher John Lennox states (Lennox 2020):

> The danger is that people are carried away with the "if it can be done, it should be done" mentally without thinking carefully through the potential ethical problems. However, it must be said that ethical issues are now rapidly rising in importance on the agenda of leading players in the field of the AI world. The big question to be faced is: How can an ethical dimension be built into an algorithm that is itself devoid of heart, soul, and mind?

In conclusion, we strongly recommend that the development and deployment of healthcare AI in Africa be ethically structured in a manner that not only aligns with the goals and values of medicine but is consistent with culturally appropriate norms that maximize benefits, minimize harms, and support the overall well-being of individuals and communities in the continent. African societies' distinct cultural, political, and economic contexts necessitate an appropriate ethical framework that should address data colonialism, informed consent, data privacy, ownership, and bias, ensuring that AI technologies are developed and used in ways that respect and reflect African values and norms. Ultimately, the goal is to promote responsible AI practices that foster a more equitable and just healthcare system.

**References**
Abebe, Rediet, Kehinde Aruleba, Abeba Birhane, Sara Kingsley, George Obaido, Sekou L. Remy, and Swathi Sadagopan. 2021. "Narratives and Counternarratives on Data Sharing in Africa." In *Proceedings of the 2021 ACM Conference on Fairness, Accountability, and Transparency*, 329–41. https://doi.org/10.1145/3442188.3445897.
Ade-Ibijola, Abejide, and Chinedu Okonkwo. 2023. "Artificial Intelligence in Africa: Emerging Challenges." In *Responsible AI in Africa: Challenges and Opportunities*, edited by Damian Okaibedi Eke, Kutoma Wakunuma, and Simisola Akintoye, 101–17. Cham: Springer International Publishing. https://doi.org/10.1007/978-3-031-08215-3_5.
Ahmed, Shamira, and Araba Sey. 2020. "An African Perspective on Gender and Artificial Intelligence Needs African Data and Research."
Al-Badi, Ali, Ali Tarhini, and Asharul Islam Khan. 2018. "Exploring Big Data Governance Frameworks." *Procedia Computer Science*, The 9th International Conference on Emerging Ubiquitous Systems and Pervasive Networks (EUSPN-2018) / The 8th International Conference on Current and Future Trends of Information and Communication Technologies in Healthcare (ICTH-2018) / Affiliated Workshops, 141 (January): 271–77. https://doi.org/10.1016/j.procs.2018.10.181.
ALT Advisory. 2022. *AI Governance in Africa*.
Appiah, Richard. 2021. "Gurus and Griots: Revisiting the Research Informed Consent Process in Rural African Contexts." *BMC Medical Ethics* 22 (1): 98. https://doi.org/10.1186/s12910-021-00659-7.
Astromskė, Kristina, Eimantas Peičius, and Paulius Astromskis. 2021. "Ethical and Legal Challenges of Informed Consent Applying Artificial Intelligence in Medical Diagnostic Consultations." *AI & SOCIETY* 36: 509–20.
Ayana, Gelan, Kokeb Dese, Hundessa Daba, Bruce Mellado, Kingsley Badu, Edmund Ilimoan Yamba, Sylvain Landry Faye, et al. 2023. "Decolonizing Global AI Governance: Assessment of the State of Decolonized AI Governance in Sub-Saharan Africa." SSRN Scholarly Paper. Rochester, NY. https://doi.org/10.2139/ssrn.4652444.
Bajwa, Junaid, Usman Munir, Aditya Nori, and Bryan Williams. 2021. "Artificial Intelligence in Healthcare: Transforming the Practice of Medicine." *Future Healthcare Journal* 8 (2): e188–94. https://doi.org/10.7861/fhj.2021-0095.

Bandura, Albert. 1999. "Moral Disengagement in the Perpetration of Inhumanities." *Personality and Social Psychology Review* 3 (3): 193–209. https://doi.org/10.1207/s15327957pspr0303_3.
Bansal, Ms. Aayushi, Dr. Rewa Sharma, and Dr. Mamta Kathuria. 2022. "A Systematic Review on Data Scarcity Problem in Deep Learning: Solution and Applications." *ACM Computing Surveys* 54 (10s): 208:1-208:29. https://doi.org/10.1145/3502287.
Barocas, Solon, Moritz Hardt, and Arvind Narayanan. 2023. *Fairness and Machine Learning: Limitations and Opportunities*. MIT Press.
Barrainkua, Ainhize, Paula Gordaliza, Jose A. Lozano, and Novi Quadrianto. 2022. "A Survey on Preserving Fairness Guarantees in Changing Environments." arXiv. https://doi.org/10.48550/arXiv.2211.07530.
Birhane, Abeba. 2020. "Algorithmic Colonization of Africa." *SCRIPTed* 17 (2): 389–409. https://doi.org/10.2966/scrip.170220.389.
Bull, Susan, Phaik Yeong Cheah, Spencer Denny, Irene Jao, Vicki Marsh, Laura Merson, Neena Shah More, et al. 2015. "Best Practices for Ethical Sharing of Individual-Level Health Research Data From Low- and Middle-Income Settings." *Journal of Empirical Research on Human Research Ethics* 10 (3): 302–13. https://doi.org/10.1177/1556264615594606.
Bull, Susan, Nia Roberts, and Michael Parker. 2015. "Views of Ethical Best Practices in Sharing Individual-Level Data From Medical and Public Health Research: A Systematic Scoping Review." *Journal of Empirical Research on Human Research Ethics* 10 (3): 225–38. https://doi.org/10.1177/1556264615594767.
Carey, Alycia N., and Xintao Wu. 2023. "The Statistical Fairness Field Guide: Perspectives from Social and Formal Sciences." *AI and Ethics* 3 (1): 1–23. https://doi.org/10.1007/s43681-022-00183-3.
Cary, Michael P., Anna Zink, Sijia Wei, Andrew Olson, Mengying Yan, Rashaud Senior, Sophia Bessias, et al. 2023. "Mitigating Racial And Ethnic Bias And Advancing Health Equity In Clinical Algorithms: A Scoping Review." *Health Affairs* 42 (10): 1359–68. https://doi.org/10.1377/hlthaff.2023.00553.
Cath, Corinne. 2018. "Governing Artificial Intelligence: Ethical, Legal and Technical Opportunities and Challenges." *Philosophical Transactions of the Royal Society A: Mathematical, Physical and Engineering Sciences* 376 (2133): 20180080. https://doi.org/10.1098/rsta.2018.0080.
Char, Danton S., Michael D. Abràmoff, and Chris Feudtner. 2020. "Identifying Ethical Considerations for Machine Learning Healthcare Applications." *The American Journal of Bioethics* 20 (11): 7–17. https://doi.org/10.1080/15265161.2020.1819469.
Cohen, I. Glenn. 2018. "Is There a Duty to Share Healthcare Data?" In *Big Data, Health Law, and Bioethics*, edited by Effy Vayena, Holly Fernandez Lynch, I. Glenn Cohen, and Urs Gasser, 209–22. Cambridge: Cambridge University Press. https://doi.org/10.1017/9781108147972.020.
Cohen, I. Glenn, Ruben Amarasingham, Anand Shah, Bin Xie, and Bernard Lo. 2014. "The Legal and Ethical Concerns That Arise from Using Complex Predictive Analytics in Health Care." *Health Affairs* 33 (7): 1139–47.
Coleman, Danielle. 2019. "Digital Colonialism: The 21st Century Scramble for Africa through the Extraction and Control of User Data and the Limitations of Data Protection Laws." *Michigan Journal of Race and Law* 24 (2): 417–39. https://doi.org/10.36643/mjrl.24.2.digital.
Davenport, Thomas, and Ravi Kalakota. 2019. "The Potential for Artificial Intelligence in Healthcare." *Future Healthcare Journal* 6 (2): 94–98. https://doi.org/10.7861/futurehosp.6-2-94.

Dawkins, Bryony, Charlotte Renwick, Tim Ensor, Bethany Shinkins, David Jayne, and David Meads. 2021. "What Factors Affect Patients' Ability to Access Healthcare? An Overview of Systematic Reviews." *Tropical Medicine & International Health* 26 (10): 1177–88. https://doi.org/10.1111/tmi.13651.

Denny, Spencer G., Blessing Silaigwana, Douglas Wassenaar, Susan Bull, and Michael Parker. 2015. "Developing Ethical Practices for Public Health Research Data Sharing in South Africa: The Views and Experiences From a Diverse Sample of Research Stakeholders." *Journal of Empirical Research on Human Research Ethics* 10 (3): 290–301. https://doi.org/10.1177/1556264615592386.

Dwork, Cynthia. 2006. "Differential Privacy." In *International Colloquium on Automata, Languages, and Programming*, 1–12. Springer.

European Health Telematics Association. 2012. *ETHICAL Principles for eHealth: Conclusions from the Consultation of the Ethics Experts around the Globe. A Briefing Paper*.

Ewuoso, C., and S. Hall. 2019. "Core Aspects of Ubuntu : A Systematic Review." *South African Journal of Bioethics and Law* 12 (2): 93–103. https://doi.org/10.7196/SAJBL.2019.v12i2.679.

Felder, Ryan Marshall. 2021. "Coming to Terms with the Black Box Problem: How to Justify AI Systems in Health Care." *Hastings Center Report* 51 (4): 38–45.

Finlayson Samuel G., Subbaswamy Adarsh, Singh Karandeep, Bowers John, Kupke Annabel, Zittrain Jonathan, Kohane Isaac S., and Saria Suchi. 2021. "The Clinician and Dataset Shift in Artificial Intelligence." *New England Journal of Medicine* 385 (3): 283–86. https://doi.org/10.1056/NEJMc2104626.

Fletcher, Richard Ribón, Audace Nakeshimana, and Olusubomi Olubeko. 2021. "Addressing Fairness, Bias, and Appropriate Use of Artificial Intelligence and Machine Learning in Global Health." *Frontiers in Artificial Intelligence* 3. https://www.frontiersin.org/articles/10.3389/frai.2020.561802.

Floridi, Luciano, and Josh Cowls. 2019. "A Unified Framework of Five Principles for AI in Society." *Harvard Data Science Review* 1 (1). https://doi.org/10.1162/99608f92.8cd550d1.

Frimpong-Mansoh, Augustine. 2008. "Culture and Voluntary Informed Consent in African Health Care Systems." *Developing World Bioethics* 8 (2): 104–14.

Gajane, Pratik, and Mykola Pechenizkiy. 2017. "On Formalizing Fairness in Prediction with Machine Learning." *arXiv Preprint arXiv:1710.03184*.

Gaonkar, Bilwaj, Kirstin Cook, and Luke Macyszyn. 2020. "Ethical Issues Arising Due to Bias in Training AI Algorithms in Healthcare and Data Sharing as a Potential Solution." *The AI Ethics Journal* 1 (1).

Gerke, Sara, Timo Minssen, and Glenn Cohen. 2020. "Ethical and Legal Challenges of Artificial Intelligence-Driven Healthcare." In *Artificial Intelligence in Healthcare*, 295–336. Elsevier.

Gilliam, Nikysha D. 2021. "Senegalese Parent, Family, and Community Engagement in Education: An Ubuntu-Inspired Inquiry." Loyola Marymount University.

Godwin Olaoye, Edwin Frank, and Ayuns Luz. 2024. "Discrimination and Stigma as Social Determinants of Depression."

Jobin, Anna, Marcello Ienca, and Effy Vayena. 2019. "The Global Landscape of AI Ethics Guidelines." *Nature Machine Intelligence* 1 (9): 389–99. https://doi.org/10.1038/s42256-019-0088-2.

Kasenene, P. 2000. "African Medical Ethics: African Ethical Theory and the Four Principles." In *Cross-Cultural Perspectives in Medical Ethics*. Jones & Bartlett Learning.

Kaur, Amandeep, Ruchi Garg, and Poonam Gupta. 2021. "Challenges Facing AI and Big Data for Resource-Poor Healthcare System." In *2021 Second International Conference on Electronics and Sustainable Communication Systems (ICESC)*, 1426–33. IEEE.

Khanna, Narendra N., Mahesh A. Maindarkar, Vijay Viswanathan, Jose Fernandes E Fernandes, Sudip Paul, Mrinalini Bhagawati, Puneet Ahluwalia, et al. 2022. "Economics of Artificial Intelligence in Healthcare: Diagnosis vs. Treatment." *Healthcare* 10 (12): 2493. https://doi.org/10.3390/healthcare10122493.

Kim, Tae Wan, and Bryan R. Routledge. 2022. "Why a Right to an Explanation of Algorithmic Decision-Making Should Exist: A Trust-Based Approach." *Business Ethics Quarterly* 32 (1): 75–102.

Kohn, Margaret, and Kavita Reddy. 2006. "Colonialism."

Lennox, John C. 2020. *2084: Artificial Intelligence and the Future of Humanity*. Zondervan.

Liu, Xiaoxuan, Livia Faes, Aditya U Kale, Siegfried K Wagner, Dun Jack Fu, Alice Bruynseels, Thushika Mahendiran, et al. 2019. "A Comparison of Deep Learning Performance against Health-Care Professionals in Detecting Diseases from Medical Imaging: A Systematic Review and Meta-Analysis." *The Lancet Digital Health* 1 (6): e271–97. https://doi.org/10.1016/S2589-7500(19)30123-2.

Lords, House Of. 2018. "AI in the UK: Ready, Willing and Able?" *Retrieved August* 13: 2021.

Martinez-Martin, Nicole, Zelun Luo, Amit Kaushal, Ehsan Adeli, Albert Haque, Sara S. Kelly, Sarah Wieten, et al. 2021. "Ethical Issues in Using Ambient Intelligence in Health-Care Settings." *The Lancet Digital Health* 3 (2): e115–23. https://doi.org/10.1016/S2589-7500(20)30275-2.

Mbiti, John S. 1990. *African Religions & Philosophy*. Heinemann.

Mbunge, Elliot, and John Batani. 2023. "Application of Deep Learning and Machine Learning Models to Improve Healthcare in Sub-Saharan Africa: Emerging Opportunities, Trends and Implications." *Telematics and Informatics Reports* 11 (September): 100097. https://doi.org/10.1016/j.teler.2023.100097.

McMahan, Brendan, Eider Moore, Daniel Ramage, Seth Hampson, and Blaise Aguera y Arcas. 2017. "Communication-Efficient Learning of Deep Networks from Decentralized Data." In *Artificial Intelligence and Statistics*, 1273–82. PMLR.

Mehrabi, Ninareh, Fred Morstatter, Nripsuta Saxena, Kristina Lerman, and Aram Galstyan. 2021. "A Survey on Bias and Fairness in Machine Learning." *ACM Computing Surveys* 54 (6): 115:1-115:35. https://doi.org/10.1145/3457607.

Metz, Thaddeus, and Joseph BR Gaie. 2010. "The African Ethic of Ubuntu/Botho: Implications for Research on Morality." *Journal of Moral Education* 39 (3): 273–90.

Micheli, Marina, Marisa Ponti, Max Craglia, and Anna Berti Suman. 2020. "Emerging Models of Data Governance in the Age of Datafication." *Big Data & Society* 7 (2): 2053951720948087. https://doi.org/10.1177/2053951720948087.

Moodley, Keymanthri, and Stuart Rennie. 2023. "The Many Faces of the Big Data Revolution in Health for Sub-Saharan Africa." *South African Journal of Science* 119 (5–6): 1–3. https://doi.org/10.17159/sajs.2023/16158.

Morley, Jessica, Caio C. V. Machado, Christopher Burr, Josh Cowls, Indra Joshi, Mariarosaria Taddeo, and Luciano Floridi. 2020. "The Ethics of AI in Health Care: A Mapping Review." *Social Science & Medicine* 260 (September): 113172. https://doi.org/10.1016/j.socscimed.2020.113172.

Muralidharan, V., A. Burgart, R. Daneshjou, and S. Rose. 2023. "Recommendations for the Use of Pediatric Data in Artificial Intelligence and Machine Learning ACCEPT-AI." *Npj Digital Medicine* 6 (1): 1–6. https://doi.org/10.1038/s41746-023-00898-5.

Murphy, Kathleen, Erica Di Ruggiero, Ross Upshur, Donald J. Willison, Neha Malhotra, Jia Ce Cai, Nakul Malhotra, Vincci Lui, and Jennifer Gibson. 2021. "Artificial Intelligence for Good Health: A Scoping Review of the Ethics Literature." *BMC Medical Ethics* 22: 1–17.

NetHope. 2020. "AI Suitability Framework For Nonprofits." 2020. https://nethope.app.box.com/s/ot0dhatbx0ddwplux2lo924mtk1p0cew.

Nienaber McKay, Annelize G, Dirk Brand, Marietjie Botes, Nezerith Cengiz, and Marno Swart. 2024. "The Regulation of Health Data Sharing in Africa: A Comparative Study." *Journal of Law and the Biosciences* 11 (1): lsad035. https://doi.org/10.1093/jlb/lsad035.

Norori, Natalia, Qiyang Hu, Florence Marcelle Aellen, Francesca Dalia Faraci, and Athina Tzovara. 2021. "Addressing Bias in Big Data and AI for Health Care: A Call for Open Science." *Patterns* 2 (10).

Nussbaum, Barbara. 2003. "Ubuntu: Reflections of a South African on Our Common Humanity." *Reflections* 4 (4): 21–26.

Oleribe, Obinna O, Jenny Momoh, Benjamin SC Uzochukwu, Francisco Mbofana, Akin Adebiyi, Thomas Barbera, Roger Williams, and Simon D Taylor-Robinson. 2019. "Identifying Key Challenges Facing Healthcare Systems In Africa And Potential Solutions." *International Journal of General Medicine* 12 (November): 395–403. https://doi.org/10.2147/IJGM.S223882.

Petrozzino, Catherine. 2021. "Who Pays for Ethical Debt in AI?" *AI and Ethics* 1 (3): 205–8.

Roppelt, Julia Stefanie, Dominik K. Kanbach, and Sascha Kraus. 2024. "Artificial Intelligence in Healthcare Institutions: A Systematic Literature Review on Influencing Factors." *Technology in Society* 76 (March): 102443. https://doi.org/10.1016/j.techsoc.2023.102443.

Sahni, Nikhil, George Stein, Rodney Zemmel, and David M. Cutler. 2023. "The Potential Impact of Artificial Intelligence on Healthcare Spending." Working Paper. Working Paper Series. National Bureau of Economic Research. https://doi.org/10.3386/w30857.

Schiff, Daniel, and Jason Borenstein. 2019. "How Should Clinicians Communicate with Patients about the Roles of Artificially Intelligent Team Members?" *AMA Journal of Ethics* 21 (2): 138–45.

Schrag, Brian. 2006. "Research with Groups: Group Rights, Group Consent, and Collaborative Research: Commentary on Protecting the Navajo People through Tribal Regulation of Research." *Science and Engineering Ethics* 12 (3): 511–21. https://doi.org/10.1007/s11948-006-0049-0.

Schwalbe, Nina, and Brian Wahl. 2020. "Artificial Intelligence and the Future of Global Health." *The Lancet* 395 (10236): 1579–86. https://doi.org/10.1016/S0140-6736(20)30226-9.

Schwalbe, Nina, Brian Wahl, Jingyi Song, and Susanna Lehtimaki. 2020. "Data Sharing and Global Public Health: Defining What We Mean by Data." *Frontiers in Digital Health* 2 (December). https://doi.org/10.3389/fdgth.2020.612339.

Shorten, Connor, and Taghi M. Khoshgoftaar. 2019. "A Survey on Image Data Augmentation for Deep Learning." *Journal of Big Data* 6 (1): 1–48. https://doi.org/10.1186/s40537-019-0197-0.

Siala, Haytham, and Yichuan Wang. 2022. "SHIFTing Artificial Intelligence to Be Responsible in Healthcare: A Systematic Review." *Social Science & Medicine* 296: 114782.

Sugarman, J., B. Popkin, J. Fortney, and R. Rivera. 2001. "International Perspectives on Protecting Human Research Subjects." *Ethical and Policy Issues in International Research: Clinical Trials in Developing Countries* 2: E1-11.

Tindana, Paulina, Sassy Molyneux, Susan Bull, and Michael Parker. 2019. "'It Is an Entrustment': Broad Consent for Genomic Research and Biobanks in sub-Saharan Africa." *Developing World Bioethics* 19 (1): 9–17.

Torkzadehmahani, Reihaneh, Reza Nasirigerdeh, David B. Blumenthal, Tim Kacprowski, Markus List, Julian Matschinske, Julian Spaeth, Nina Kerstin Wenke, and Jan Baumbach. 2022. "Privacy-Preserving Artificial Intelligence Techniques in Biomedicine." *Methods of Information in Medicine* 61 (Suppl 1): e12–27. https://doi.org/10.1055/s-0041-1740630.

Townsend, B. A., and R. E. Scott. 2019. "The Development of Ethical Guidelines for Telemedicine in South Africa." *South African Journal of Bioethics and Law* 12 (1): 19–26. https://doi.org/10.7196/SAJBL.2019.v12i1.662.

Townsend, Beverley Alice, Irvine Sihlahla, Meshandren Naidoo, Shiniel Naidoo, Dusty-Lee Donnelly, and Donrich Willem Thaldar. 2023. "Mapping the Regulatory Landscape of AI in Healthcare in Africa." *Frontiers in Pharmacology* 14 (August): 1214422. https://doi.org/10.3389/fphar.2023.1214422.

UNESCO, C. 2021. *Recommendation on the Ethics of Artificial Intelligence*. UNESCO Geneva, Switzerland.

United Nations Conference on Trade and Development, R. M. 2021. *Data Protection and Privacy Legislation Worldwide*. Geneva.

Vibbi, Leonard Francis. 2024. "Poor Data Quality in Sub-Saharan Africa and Implications on Ethical AI Development." In *Improving Technology Through Ethics*, edited by Simona Chiodo, David Kaiser, Julie Shah, and Paolo Volonté, 83–92. Cham: Springer Nature Switzerland. https://doi.org/10.1007/978-3-031-52962-7_7.

Waghid, Yusef. 2013. *African Philosophy of Education Reconsidered*. 0 ed. Routledge. https://doi.org/10.4324/9780203538166.

Wahl, Brian, Aline Cossy-Gantner, Stefan Germann, and Nina R. Schwalbe. 2018. "Artificial Intelligence (AI) and Global Health: How Can AI Contribute to Health in Resource-Poor Settings?" *BMJ Global Health* 3 (4): e000798. https://doi.org/10.1136/bmjgh-2018-000798.

Wareham, Christopher Simon. 2017. "Partiality and Distributive Justice in African Bioethics." *Theoretical Medicine and Bioethics* 38 (2): 127–44. https://doi.org/10.1007/s11017-017-9401-4.

World Health Organization. 2020. "World Health Statistics 2020."